\documentstyle[prd,floats,aps,epsfig,twocolumn]{revtex}

\def\Missing#1#2{{\mbox{$#1\kern-0.57em\raise0.19ex\hbox{/}_{#2}$}}}

\def\vMissing#1#2{\ifmmode
            \vec{#1}\kern-0.57em\raise.19ex\hbox{/}_{#2}
         \else
            {{\mbox{$\vec{#1}\kern-0.57em\raise.19ex\hbox{/}_{#2}$}}}
         \fi}

\def\D0{D\O }

\def\ETmiss{\mbox{${\hbox{$E$\kern-0.5em\lower-.1ex\hbox{/}\kern+0.15em}}_T$ }}

\def\err#1#2#3 {{\it Erratum} {\bf#1},{\ #2} (19#3)}
\def\ib#1#2#3 {{\it ibid.} {\bf#1},{\ #2} (19#3)}
\def\nc#1#2#3 {Nuovo Cim. {\bf#1} ,#2(19#3)}
\def\nim#1#2#3 {Nucl. Instr. Meth. {\bf#1},{\ #2} (19#3)}
\def\np#1#2#3 {Nucl. Phys. {\bf#1},{\ #2} (19#3)}
\def\pl#1#2#3 {Phys. Lett. {\bf#1},{\ #2} (19#3)}
\def\prev#1#2#3 {Phys. Rev. {\bf#1},{\ #2} (19#3)}
\def\prl#1#2#3 {Phys. Rev. Lett. {\bf#1},{\ #2} (19#3)}
\def\rmp#1#2#3 {Rev. Mod. Phys. {\bf#1},{\ #2} (19#3)}
\def\zp#1#2#3 {Zeit. Phys. {\bf#1},{\ #2} (19#3)}     

\newcommand{\gapprox}{\raisebox{-0.7ex}{$\stackrel {\textstyle>}{\sim}$}}

\newcommand{\bls}   {\mbox{$\tilde{t} \rightarrow  b \ell \snu        $}}
\newcommand{\bc}    {\mbox{$\tilde{t} \rightarrow  b \ca  $}}
\newcommand{\blstau}{\mbox{$\tilde{t} \rightarrow  b \tau \snu_{\tau} $}}

\newcommand{\etm}{\mbox{$E_T \! \! \! \! \! \! \!/ \ $}}

\newcommand{\bnu}{\mbox{$\bar{\nu}$}}

\newcommand{\oa}{\mbox{$\tilde{\chi}^{0}_1$}}
\newcommand{\mca}{m_{{\tilde{\chi}^+_1}}}
\newcommand{\ca}{\mbox{$\tilde{\chi}^+_1$}}

\newcommand{\snu}{\mbox{$\tilde{\nu}$}}
\newcommand{\msnu}{m_{\tilde{\nu}}}
\newcommand{\mst}{m_{\tilde{t}}}

\newcommand{\stt}{\mbox{$\tilde{t}$}}
\newcommand{\stb}{\mbox{$\bar{\tilde{t}}$}}

\newcommand{\arrow}{\mbox{$\rightarrow \ $}}

\begin{document}
\lefthyphenmin=3
\righthyphenmin=3

\onecolumn
 \title{A Search for 
the Scalar Top Quark    in $p \bar{p} $ Collisions  
 at $\sqrt{s}=$ 1.8   TeV} 
%
\author{                                                                      
V.M.~Abazov,$^{23}$                                                           
B.~Abbott,$^{58}$                                                             
A.~Abdesselam,$^{11}$                                                         
M.~Abolins,$^{51}$                                                            
V.~Abramov,$^{26}$                                                            
B.S.~Acharya,$^{17}$                                                          
D.L.~Adams,$^{60}$                                                            
M.~Adams,$^{38}$                                                              
S.N.~Ahmed,$^{21}$                                                            
G.D.~Alexeev,$^{23}$
% ...
\\                                                                            
\vskip 0.30cm                                                                 
\centerline{(D\O\ Collaboration)}                                             
\vskip 0.30cm                                                                 
}                                                                             
\address{                                                                     
\centerline{$^{1}$Universidad de Buenos Aires, Buenos Aires, Argentina}       
\centerline{$^{2}$LAFEX, Centro Brasileiro de Pesquisas F{\'\i}sicas,         
                  Rio de Janeiro, Brazil}                                     
\centerline{$^{3}$Universidade do Estado do Rio de Janeiro,                   
                  Rio de Janeiro, Brazil}                                     
\centerline{$^{4}$Institute of High Energy Physics, Beijing,                  
                  People's Republic of China}                                 
\centerline{$^{5}$Universidad de los Andes, Bogot\'{a}, Colombia}             
\centerline{$^{6}$Charles University, Center for Particle Physics,            
                  Prague, Czech Republic}                                     
\centerline{$^{7}$Institute of Physics, Academy of Sciences, Center           
                  for Particle Physics, Prague, Czech Republic}               
\centerline{$^{8}$Universidad San Francisco de Quito, Quito, Ecuador}         
\centerline{$^{9}$Institut des Sciences Nucl\'eaires, IN2P3-CNRS,             
                  Universite de Grenoble 1, Grenoble, France}                 
\centerline{$^{10}$CPPM, IN2P3-CNRS, Universit\'e de la M\'editerran\'ee,     
                  Marseille, France}                                          
\centerline{$^{11}$Laboratoire de l'Acc\'el\'erateur Lin\'eaire,              
                  IN2P3-CNRS, Orsay, France}                                  
\centerline{$^{12}$LPNHE, Universit\'es Paris VI and VII, IN2P3-CNRS,         
                  Paris, France}                                              
\centerline{$^{13}$DAPNIA/Service de Physique des Particules, CEA, Saclay,    
                  France}                                                     
\centerline{$^{14}$Universit{\"a}t Mainz, Institut f{\"u}r Physik,            
                  Mainz, Germany}                                             
\centerline{$^{15}$Panjab University, Chandigarh, India}                      
\centerline{$^{16}$Delhi University, Delhi, India}                            
\centerline{$^{17}$Tata Institute of Fundamental Research, Mumbai, India}     
\centerline{$^{18}$Seoul National University, Seoul, Korea}                   
\centerline{$^{19}$CINVESTAV, Mexico City, Mexico}                            
\centerline{$^{20}$FOM-Institute NIKHEF and University of                     
                  Amsterdam/NIKHEF, Amsterdam, The Netherlands}               
\centerline{$^{21}$University of Nijmegen/NIKHEF, Nijmegen, The               
                  Netherlands}                                                
\centerline{$^{22}$Institute of Nuclear Physics, Krak\'ow, Poland}            
\centerline{$^{23}$Joint Institute for Nuclear Research, Dubna, Russia}       
\centerline{$^{24}$Institute for Theoretical and Experimental Physics,        
                   Moscow, Russia}                                            
\centerline{$^{25}$Moscow State University, Moscow, Russia}                   
\centerline{$^{26}$Institute for High Energy Physics, Protvino, Russia}       
\centerline{$^{27}$Lancaster University, Lancaster, United Kingdom}           
\centerline{$^{28}$Imperial College, London, United Kingdom}                  
\centerline{$^{29}$University of Arizona, Tucson, Arizona 85721}              
\centerline{$^{30}$Lawrence Berkeley National Laboratory and University of    
                  California, Berkeley, California 94720}                     
\centerline{$^{31}$University of California, Davis, California 95616}         
\centerline{$^{32}$California State University, Fresno, California 93740}     
\centerline{$^{33}$University of California, Irvine, California 92697}        
\centerline{$^{34}$University of California, Riverside, California 92521}     
\centerline{$^{35}$Florida State University, Tallahassee, Florida 32306}      
\centerline{$^{36}$University of Hawaii, Honolulu, Hawaii 96822}              
\centerline{$^{37}$Fermi National Accelerator Laboratory, Batavia,            
                   Illinois 60510}                                            
\centerline{$^{38}$University of Illinois at Chicago, Chicago,                
                   Illinois 60607}                                            
\centerline{$^{39}$Northern Illinois University, DeKalb, Illinois 60115}      
\centerline{$^{40}$Northwestern University, Evanston, Illinois 60208}         
\centerline{$^{41}$Indiana University, Bloomington, Indiana 47405}            
\centerline{$^{42}$University of Notre Dame, Notre Dame, Indiana 46556}       
\centerline{$^{43}$Iowa State University, Ames, Iowa 50011}                   
\centerline{$^{44}$University of Kansas, Lawrence, Kansas 66045}              
\centerline{$^{45}$Kansas State University, Manhattan, Kansas 66506}          
\centerline{$^{46}$Louisiana Tech University, Ruston, Louisiana 71272}        
\centerline{$^{47}$University of Maryland, College Park, Maryland 20742}      
\centerline{$^{48}$Boston University, Boston, Massachusetts 02215}            
\centerline{$^{49}$Northeastern University, Boston, Massachusetts 02115}      
\centerline{$^{50}$University of Michigan, Ann Arbor, Michigan 48109}         
\centerline{$^{51}$Michigan State University, East Lansing, Michigan 48824}   
\centerline{$^{52}$University of Nebraska, Lincoln, Nebraska 68588}           
\centerline{$^{53}$Columbia University, New York, New York 10027}             
\centerline{$^{54}$University of Rochester, Rochester, New York 14627}        
\centerline{$^{55}$State University of New York, Stony Brook,                 
                   New York 11794}                                            
\centerline{$^{56}$Brookhaven National Laboratory, Upton, New York 11973}     
\centerline{$^{57}$Langston University, Langston, Oklahoma 73050}             
\centerline{$^{58}$University of Oklahoma, Norman, Oklahoma 73019}            
\centerline{$^{59}$Brown University, Providence, Rhode Island 02912}          
\centerline{$^{60}$University of Texas, Arlington, Texas 76019}               
\centerline{$^{61}$Texas A\&M University, College Station, Texas 77843}       
\centerline{$^{62}$Rice University, Houston, Texas 77005}                     
\centerline{$^{63}$University of Virginia, Charlottesville, Virginia 22901}   
\centerline{$^{64}$University of Washington, Seattle, Washington 98195}       
}                                                                             

\maketitle

\vskip 10pt

\date{\today}
{
{\bf
\begin{center}
Abstract
\end{center}
}
\begin{center}
\begin{minipage}{.8\textwidth}
{\small 
We have performed a search for scalar top quark (stop) pair production in
the inclusive electron-muon-missing transverse energy 
final state, using a
sample of $p \bar{p}$ events corresponding to $108.3$ pb$^{-1}$
of  data collected with the D\O\ detector at Fermilab. The
search is done 
in the framework
of the minimal supersymmetric
standard model 
assuming that the sneutrino  
is the lightest supersymmetric particle. 
For the dominant  decays of the lightest
stop, $\bc$ and  $\bls$, no evidence for signal is found.
We derive cross-section limits as a function of stop (\stt), 
chargino (\ca), and sneutrino  (\snu)  masses.
}
\end{minipage}
\end{center}

\vskip 2.0in

\twocolumn

%
%

\normalsize

\vfill\eject

Supersymmetry (SUSY)~\cite{susy} 
provides a theoretically attractive and coherent picture of
the microscopic world
 that retains the standard model's 
successful description of the observed elementary
particles and their interactions.
A major  consequence of the realization of SUSY in nature
would be the 
existence of
additional particles (sparticles), with
quantum numbers identical to those of the
elementary particles of the standard model (SM), but with spins
differing by  a  half unit. From experimental evidence,
the sparticle  masses  also differ from those of their SM partners, i.e.,
SUSY is a broken symmetry, and  it is expected that the
mass spectrum of the sparticles has a  different pattern than that of
the SM. 
In particular, in several SUSY models, 
the large mass of the top quark ($m_t$)
induces 
a strong
mixing between the  supersymmetric partners 
of the two chirality states of the
top quark leading naturally to two physical states, 
 \stt $_1$ and \stt $_2$, of  very different mass~\cite{ellis}.
The lightest stop  quark 
\stt$_1$ (called \stt \ in this Letter) could therefore   be
significantly lighter than the other squarks
rendering it  a particularly auspicious choice 
for a direct search.

The  production of a  pair of stop
quarks (\stt \stb) at the Tevatron can proceed through
 gluon fusion or quark annihilation. 
The cross section 
for such a process depends to a large extent only on the stop     
mass $\mst$,
and is known  at next-to-leading order (NLO)
with a precision of $\pm 8\%$~\cite{prospino}.
The phenomenology of stop      
decays depends  on  the assumptions of the
SUSY model, and this analysis is done in the 
minimal supersymmetric
standard model 
(MSSM)~\cite{mssm} framework
with $R$-parity~\cite{rparity} conservation, 
implying that the lightest SUSY  particle (LSP)  is stable. 
Searches for stop      production have  already been performed 
at the Tevatron
assuming that the lightest neutralino (\oa ) is the LSP~\cite{stop-refs}.

In this Letter we also search for light stop      ($\mst < m_t$) 
production, but assume that the sneutrino ($\tilde{\nu}$)
is the LSP. 
Stop searches have been performed under these
assumptions 
at LEP~2~\cite{lep-stop} and by the CDF collaboration 
at the Tevatron~\cite{cdf-snu}  
yielding a mass limit $\mst \ \gapprox  $  $123$~GeV for the
lowest allowed sneutrino mass, $\msnu \approx 45$~GeV,
 as determined at LEP~1~\cite{pdglep1}.
Although these analyses are interpreted in the framework of the MSSM,
the results are largely model independent, depending mainly on the masses of
the stop      and its decay products.

In the stop      mass range probed by
the Tevatron, 
either the 2-body decay via a chargino, $\bc$,
 is  kinematically
allowed and thereby dominant, 
or the chargino mediating the decay is virtual and
the dominant decay mode is
$\bls$.
The three other $3$-body decays
mediated by a chargino,  $\tilde{t} \rightarrow b \nu \tilde{\ell}^+$
($\rightarrow  b \nu \ell^+ \oa$), 
  $\tilde{t} \rightarrow b W \oa $ and 
  $\tilde{t} \rightarrow b  H^+ \oa$, with  subsequent decays  $\oa \arrow
\snu \nu $, 
are disfavored~\cite{porod}.
In this Letter, the chargino is taken either as virtual with a 
propagator mass of
 $140$~GeV, or its mass  is varied between its lowest 
experimental limit ($\approx 103$~GeV~\cite{lep2-chargino}) 
and the maximum value allowed by kinematics.
The  masses of the sneutrinos of all three flavors are taken to be equal,
except when
the channel $\blstau$ is assumed to be dominant.

The experimental signature for decays of a  \stt \stb \ pair 
 consists of   two $ b$
quarks, two leptons, and missing transverse energy (\etm \ ). 
The variable $\etm \ $ represents the measured imbalance in
 transverse  energy  due to 
the two escaping sneutrinos.
The leptons can be
$e, \mu$ or $\tau$, but $\tau$ leptons are considered
only if they
decay into $e\nu \bar{\nu}$ or $\mu \nu \bar{\nu}$.
We place no requirements on the presence of jets and use 
only the  $e \mu \etm \  $ 
signature  since it has less background
than the $ee \etm \ $ or $\mu\mu \etm \ $ 
channels. 
The resulting event sample
corresponds to  $108.3$~pb$^{-1}$ of data
collected by the D\O\ experiment at Fermilab during the Run I of the Tevatron.

A detailed description of the D\O\ detector and its triggering system
can be found in Ref.~\cite{D0Detector}. The data and  pre-selection criteria 
are identical to those used in the published $t \bar{t}$ cross section
analysis for the dilepton channel~\cite{D0-dilepton}, which includes the 
selection of events containing one or more isolated electrons with
$E_T^e > 15$~GeV, one or more isolated muons with $E_T^{\mu} > 15$~GeV, 
and  $\etm \ > 20$~GeV. 
 $\etm \ $ 
is obtained from
the vector sum of the transverse energy measured in the calorimeter and 
in the muon spectrometer system.
Electrons  are required to have $| \eta_{\text{det}} | < 1.1$,
 or $1.5 < | \eta_{\text{det}}| < 2.5$,  where $\eta_{\text{det}}$ is the
pseudorapidity ($\eta$) defined with respect to the center of the detector.
Muons must satisfy  $| \eta_{\text{det}} | < 1.7$.

The dominant SM processes that provide the 
$e \mu \etm \ $   signature are, in order of decreasing importance:
i) multi-jet  processes (called ``QCD'' in the following)
with one jet misidentified as an electron
and one true muon originating from another jet (muon misidentification
has negligible effects on our final state);
ii) $Z \rightarrow \tau^+ \tau^- \arrow e\mu\nu\bnu\nu\bnu$;
iii) $WW\arrow e\mu\nu\bnu$;
iv) $t \bar{t} \arrow e\mu \nu \bnu j j$; and
v) Drell-Yan ($DY$) \arrow  $\tau^+\tau^- 
 \arrow e\mu\nu\bnu\nu\bnu$. 
The QCD background
was determined from  data, following the procedure described
in Ref.~\cite{92-93-top}. 
The other backgrounds were simulated and reconstructed 
using the full D\O\ analysis chain.


Simulation of the signal 
is based on {\sc {Pythia}}~\cite{pythia}, 
using the { {CTEQ3M}}~\cite{cteq3} parton distribution functions (PDFs),
and the standard 
 hadronization and fragmentation functions in {\sc {Pythia}}. 
{\sc{Comphep}}~\cite{comphep} is used to generate
the $2$ and  $3$-body decays of
the stop. 
%
%
Detector  simulation is performed using  the
fast D\O\ simulation/reconstruction program,
which has been   checked 
extensively on a reference sample passed through 
the full D\O\ analysis chain.
The \stt \stb\ samples were simulated for
stop      (sneutrino, chargino) masses varying between $50 \ (30,100)$ and  
$150 \ (90,170)$~GeV. 

Distributions in the kinematic   quantities
($E_T^e, E_T^{\mu}$,  \etm~) are
shown in Fig.~\ref{nocuts}(a--c). Also shown (d) 
are the  distributions for the
transverse energy of any associated  jets,
defined by a cone algorithm  and having  $E_T^{\text{jet}}>15$~GeV,  and
two additional kinematic quantities in which the signal and
background display a different response: (e) 
$\Delta_{\varphi}^{e\mu} \equiv
 | \varphi_e-\varphi_{\mu} | $,
where $\varphi_{\ell}$ is the azimuthal angle of the lepton $\ell$, and 
(f) $\Sigma_{\eta}^{e\mu} \equiv  | \eta_e+\eta_{\mu} | $.
Based on simulation studies,
two additional  criteria,
  $ 15^{\circ} < \Delta_{\varphi}^{e\mu} < 165^{\circ}$ and
    $\Sigma_{\eta}^{e\mu} < 2.0 $, were applied 
to improve the  signal to background  ratio in the final sample.
\begin{figure}[t]
\centerline{\psfig{
figure=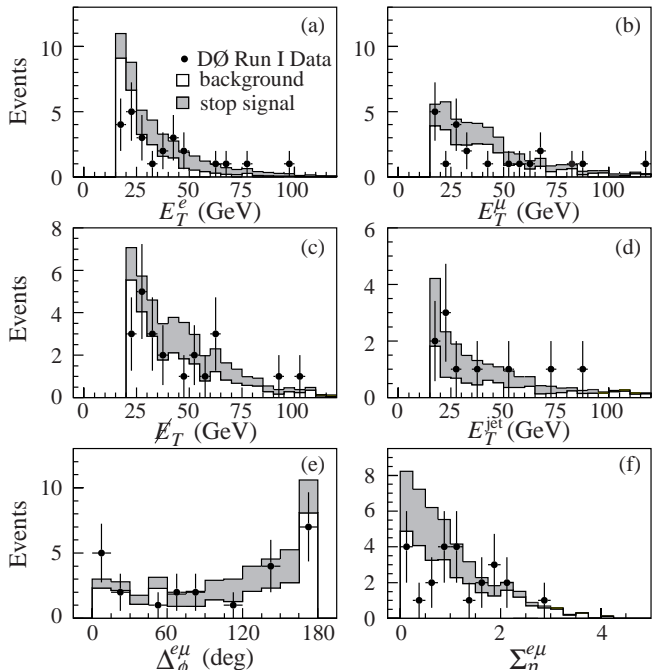,width=3.4in
}}
\caption{
Distributions after pre-selection for the total background (open histogram),
the sum of the total background and the expected stop      signal for 
$\mst$ $(\msnu) = 120$ $(60)$~GeV
(shaded histogram) 
and the data (points) of 
(a) the  transverse energy of  the electron,
(b) the  transverse energy of  the muon,
(c) the missing transverse energy,
(d) the  transverse energy of  the jets,
(e) the difference in azimuthal angle between the two leptons, and
(f) the absolute value of the sum  in $\eta$ of the two leptons.
}
\label {nocuts}
\end{figure}

The expected cross sections for the background processes, 
the normalized  numbers of events passing the pre-selection  
and those passing the final selection are given in Table~\ref{tab:result},
and  compared to the expected 
stop      signal for $\mst$ $(\msnu) = 120$ $(60)$~GeV. 
The  efficiency for selecting 
the signal varies typically between $1\%$ and $4\%$.
The most significant sources of uncertainties on the signal
are the trigger and lepton identification efficiencies ($\approx 12\%$),
the stop      pair production cross section ($8\%$), the uncertainty
due to the PDFs ($5\%$)~\cite{beena}, 
the effect of the analysis criteria ($6\%$) and 
the luminosity ($5.3\%$), 
which combine 
to approximately $18\%$. 
This uncertainty also includes the effect of the variation of the
SUSY  parameters $\mu_{\text{susy}}$ (the higgs-higgsino mass parameter)
and $\mca$~\cite{olivier}. 
The systematic error for the background is  about $10\%$. This error 
is dominated by the uncertainty on the QCD background ($7\%$) 
and on the cross sections for the background processes (10--17\%).


\begin{table}[bth]
\begin{center}
\begin{tabular}{|l|c|c|c|}
Process & Cross section  &      Events after  &    Events after \\
        &  ($\rm{pb}$)   &   pre-selection   &  final selection \\
\hline\hline
``QCD''     &  $  -   $  & $ 15.1 \pm 1.3$  & $  6.7 \pm 0.5 $ \\
$Z \arrow \tau^+ \tau^-$ &  
               $ 1.70 $  &  $  5.3 \pm 1.0 $  & $  1.4 \pm 0.3 $ \\
$WW    $ 
            &  $ 0.69 $  &  $  4.4 \pm 0.7 $  & $  3.3 \pm 0.3 $ \\
$t \bar{t}   $ 
            &  $ 0.40 $  &  $  2.7 \pm 0.5 $  & $  2.2 \pm 0.4 $ \\
$DY  \arrow \tau^+ \tau^-$
            &  $ 0.35  $ & $  0.18  \pm 0.04 $ & $  0.04 \pm 0.02 $ \\
\hline
Total backg.
            &  $  -   $  &  $ 27.8 \pm 2.7 $  & $ 13.7 \pm 1.5 $ \\
\hline
Data      
            &  $  -   $  &  $ 24      $      & $ 10  $ \\
\hline
\stt \stb  \ 
            &  $  4.51   $  & $  17.3 \pm 3.1   $       & $ 13.2 \pm 2.3 $ \\
\end{tabular}
\caption{
Cross~sections 
for the background processes, 
the expected  numbers of simulated
events passing the pre-selection  and the final analysis criteria,
numbers of events selected in 
the $e \mu \etm \ $ data sample
and the expected  
stop      signal  assuming
$\mst$ $(\msnu) =120$ $(60)$~GeV.
}
\label{tab:result} 
\end{center}
\end{table}

The agreement between  the number of observed  events and the
expected background leads us to set
 cross-section limits on  stop
quark pair production.
The $95\%$ confidence level (C.L.) limits
are obtained using a Bayesian approach~\cite{Bert}
that takes  statistical and systematic uncertainties into account. 
Assuming that the stop      decays via a virtual chargino and
$\msnu = 50$~GeV,
any stop      mass  between $73$ and  $143$~GeV is
 excluded. 
The CDF collaboration
has also performed a search in the \bls\ channel~\cite{cdf-snu}, but
based on a different signature: 
large missing transverse energy, at least
one lepton, one jet
identified as a $b$ jet, and at least another jet.
The CDF and D\O\ 
results are
compared in Fig.~\ref{xsec.limit.cdf}. 

In the MSSM,  
when the ratio of the two 
vacuum expectation values of the Higgs fields
is large ($\tan\beta \gapprox 10$),
the $\tilde{\nu}_{\tau}$ can be substantially
lighter than the $\tilde{\nu}_{e}$ or the  $\tilde{\nu}_{\mu}$,
leading to an enhancement of the decay width for 
$\blstau$~\cite{porod,djouadi-stau}. 
In this case, 
the absence of signal provides a limit on the
 cross section in  this decay
channel, as shown in Fig.~\ref{xsec.limit.cdf}
for $m_{\tilde{\nu}_{\tau}}=50$~GeV. 
\begin{figure}[tbhp]
\centerline{\psfig{
figure=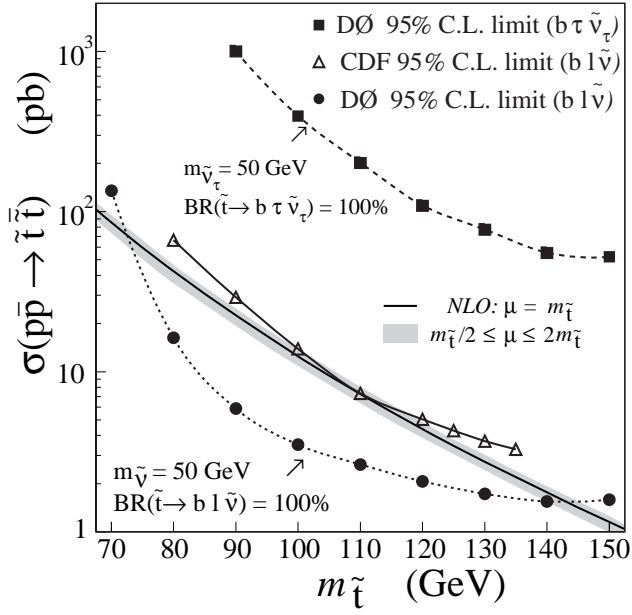,width=3.4in
}}
\caption{
Cross-section limit as a function of
$\mst$ for  $\msnu=50$~GeV.
The $\bls$ results of this analysis are compared 
to those of  CDF
 and to
 the expected NLO  cross section for three different choices of 
 factorization scale $\mu$. The renormalization scale is taken to be equal to
$\mu$. Also shown
is the limit obtained in  the $\blstau$ channel for  
$m_{\tilde{\nu}_{\tau}}=50$~GeV.
}
\label{xsec.limit.cdf}
\end{figure}

Again assuming  lepton universality, 
more \stt \stb\ production limits 
are shown in Fig.~\ref{xsec.limit.sne} for different $\msnu$ values.
For a fixed value of  $\mst$, the cross-section limit becomes
stronger with  decreasing sneutrino mass, although
the difference between limits obtained for different $\msnu $ decreases
for  high $\mst$.
For  $\msnu$ up to  $85$~GeV, and
for certain values of $\mst$,
these  are  below the
expected MSSM cross sections. 

\begin{figure}[htbp]
\centerline{\psfig{
figure=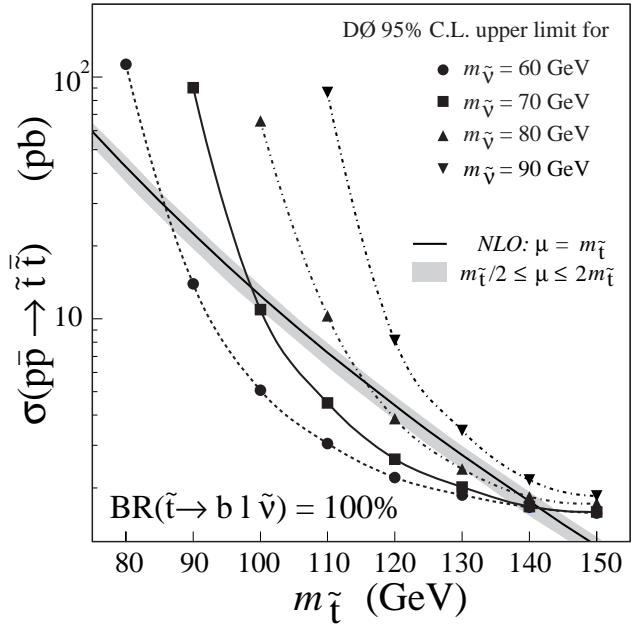,width=3.4in
}}
\caption{
Limits on the stop      pair production cross section as a function of
 $\mst$, for  $\msnu=60, 70, 80$ and 
 $90$~GeV. 
 These limits are compared to
 the expected NLO  cross section for three different choices of 
 factorization scale  $\mu$. 
}
\label {xsec.limit.sne}
\end{figure}

The resulting exclusion contour 
in the                          ($\mst$,$m_{\tilde{\nu}}$) plane
is displayed in Fig.~\ref{sne.limit},
and compared to those
obtained 
by CDF~\cite{cdf-snu},
 LEP~1, and most recently at LEP~2~\cite{lep2-stop}. 
The present analysis places limits at significantly higher $\mst$ 
compared to these results. This is mainly because of  the higher
center of mass energy of the Tevatron compared to LEP,
and of the choice of a            more sensitive 
signature compared to CDF.
For $\msnu  = 45$~GeV, the excluded region 
extends up 
to a  scalar top  
mass of $144$~GeV, to be compared to approximately 
$123$ ($98$)~GeV for CDF (LEP~2).

\begin{figure}[htbp]
\centerline{\psfig{
figure=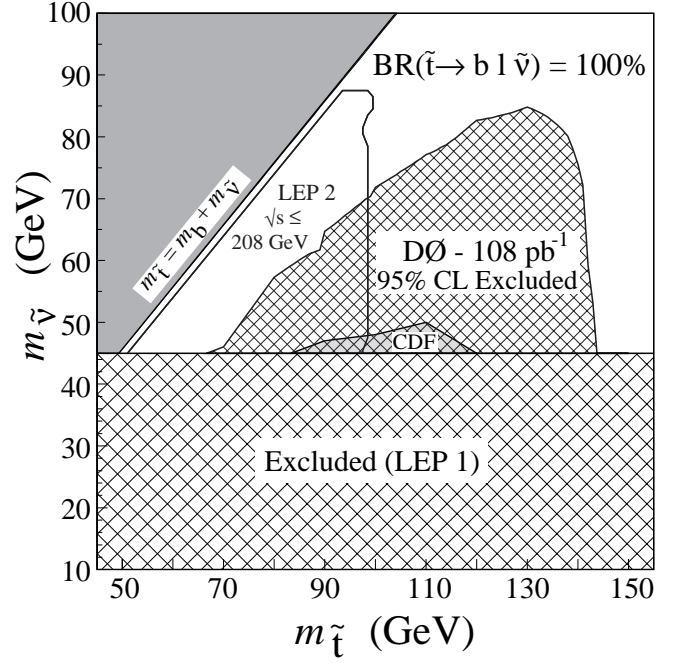,width=3.4in
}}
\caption{
Excluded regions  in the 
 ($\mst,\msnu$) plane for the $\bls$ decay channel
in the MSSM.
The results of this analysis
 (labelled D\O\ $108$ pb$^{-1}$)
  are compared to the exclusion limits 
obtained in the \bls\ decay channel  at the Tevatron (CDF), 
and at LEP~2. Also shown is 
the sneutrino mass limit obtained at LEP~1.
}
\label {sne.limit}
\end{figure}

The $2$-body decay into a $b$ quark and a real chargino, $\bc$,  was
simulated for $\mca$
 between $100$ and $140$~GeV, and the $\ca$
was assumed to 
decay only into $\ell \snu$, leading to the same final state as $\bls$.
Figure~\ref{chi.limit}
\begin{figure}[htbp]
\centerline{\psfig{
figure=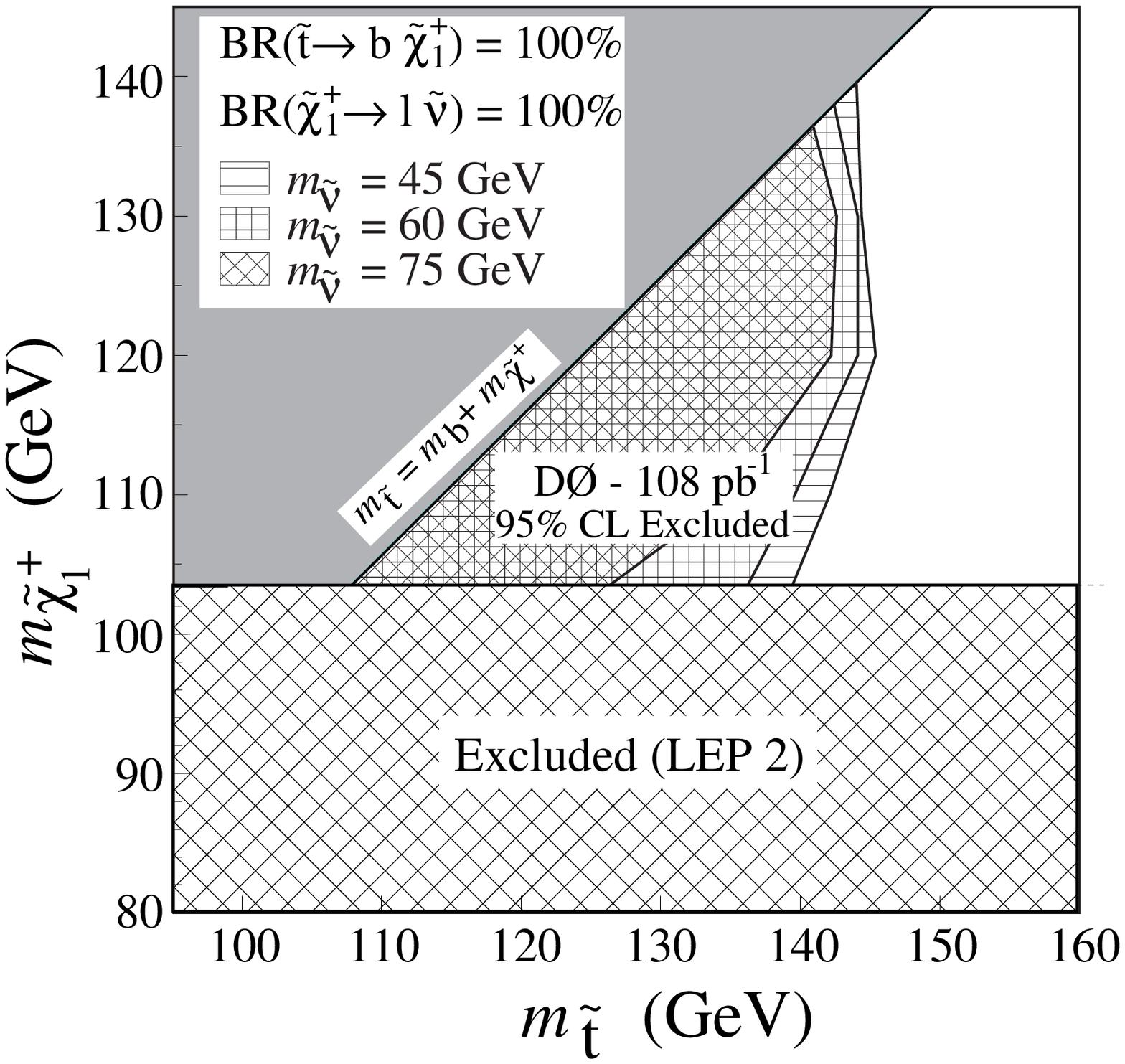,width=3.6in
}}
\caption{
Excluded regions in the 
 ($\mst,\mca$) plane for the $\tilde{t} \arrow b \ca$ 
decay channel in the MSSM, for $\msnu=45, 60$ and 75~GeV. 
These results
are compared to the exclusion limit
obtained at LEP~2. 
}
\label {chi.limit}
\end{figure}
shows 
exclusion contours  as a function of $\mst$ and
$\mca$, 
 assuming $\msnu=45, 60$ or 75~GeV. 
They are compared to the exclusion limit
obtained at LEP~2  assuming
unification of the gaugino masses and decay of the chargino via a 
$W^*$~\cite{lep2-chargino}.

In conclusion,  our analysis that assumes
the $\snu$ to be the LSP places new limits
on the stop      mass.
Assuming lepton universality and a virtual intermediary chargino,
the excluded region at $95\%$ C.L. extends up 
to a  scalar top  
mass of $144$~(130)~GeV for $\msnu = 45$~(85)~GeV.\\

%
We thank the staffs at Fermilab and collaborating institutions, 
and acknowledge support from the 
Department of Energy and National Science Foundation (USA),  
Commissariat  \` a L'Energie Atomique and 
CNRS/Institut National de Physique Nucl\'eaire et 
de Physique des Particules (France), 
Ministry for Science and Technology and Ministry for Atomic 
   Energy (Russia),
CAPES and CNPq (Brazil),
Departments of Atomic Energy and Science and Education (India),
Colciencias (Colombia),
CONACyT (Mexico),
Ministry of Education and KOSEF (Korea),
CONICET and UBACyT (Argentina),
The Foundation for Fundamental Research on Matter (The Netherlands),
PPARC (United Kingdom),
Ministry of Education (Czech Republic),
and the A.P.~Sloan Foundation.
%



\begin{references}
\vspace*{1cm}
%
\bibitem[*]{lehner}
Visitor from University of Zurich, Zurich, Switzerland.
%
\vskip 0.25cm

\bibitem{susy} 
Y. Golfand and E. Likthman, JETP Lett. {\bf 13}, 323 (1971);
D. Volkov and V. Akulov, Phys. Lett. B {\bf 46}, 109 (1973);
J. Wess and B. Zumino, Nucl. Phys. B {\bf 70}, 31 (1974);
{\sl ibid.} {\bf 78}, 1 (1974).

\bibitem {ellis} J. Ellis and S. Rudaz, Phys. Lett. B {\bf 128}, 248 (1983);
M. Drees and K. Hikasa, Phys. Lett. B {\bf 252}, 127 (1990).

\bibitem{prospino}
 W. Beenakker, R. Hopker, M. Spira, hep-ph/9611232.


\bibitem{mssm} 
H.~P.~Nilles,
Phys.\ Rep.\  {\bf 110}, 1 (1984);
H.~E.~Haber and G.~L.~Kane,
Phys.\ Rep.\  {\bf 117}, 75 (1985).

\bibitem{rparity} 
P.~Fayet,
Phys.\ Lett.\ B {\bf 69}, 489 (1977);
G.~R.~Farrar and P.~Fayet,
Phys.\ Lett.\ B {\bf 76}, 575 (1978).

%

\bibitem {stop-refs}
D\O\ Collaboration, S. Abachi {\sl et al.}, 
Phys. Rev. Lett. {\bf 76}, 2222 (1996);
D\O\ Collab., S. Abachi {\sl et al.}, 
Phys. Rev. D {\bf 57} 589 (1998);
CDF Collab., T. Affolder {\sl et al.}, 
Phys. Rev. Lett. {\bf 84}, 5704 (2000);
{\sl ibid.} {\bf 84}, 5273 (2000).

\bibitem {lep-stop}
ALEPH Collaboration, R.~Barate {\sl et al.}, 
Phys. Lett. B {\bf 469}, 303 (1999);
DELPHI Collab., P.~Abreu {\sl et al.},
Phys. Lett. B {\bf 496}, 59 (2000);
L3 Collab., M.~Acciarri {\sl et al.},
Phys. Lett. B {\bf 471}, 308 (1999);
OPAL Collab., G.~Abbiendi {\sl et al.},
Phys. Lett. B {\bf 456}, 95 (1999).

\bibitem{cdf-snu}
CDF Collaboration, T. Affolder {\sl et al.}, 
Phys. Rev. Lett. {\bf 84}, 5273 (2000).

\bibitem{pdglep1}
Particle Data Group, D. E. Groom {\sl et al.}, 
Eur. Phys. J.  C {\bf 15}, 1  (2000), and 2001 off-year partial update for
the 2002 edition), {\tt http://pdg.lbl.gov/}.


\bibitem{porod} 
W. Porod, Phys. Rev. D {\bf 59}, 095009 (1999).

\bibitem{lep2-chargino}
LEP SUSY Working Group, ALEPH, DELPHI, L3 and OPAL Collaborations, 
LEPSUSYWG/01-03.1 (2001),
{\tt http://lepsusy.web.cern.ch/lepsusy/}.

\bibitem{D0Detector}
{D\O\ Collaboration}, S.~Abachi {\sl et al}.,
Nucl. Instr. and Methods~{\bf A338}, 185 (1994).

\bibitem{D0-dilepton}
{D\O\ Collaboration}, S.~Abachi {\sl et al}., 
{Phys. Rev. Lett.} 
{\bf 79}, 1203
{(1997)}.

\bibitem{92-93-top}
D\O\ Collaboration, S. Abachi {\sl et al.}, 
Phys. Rev. D {\bf 52}, 4877 (1995).

\bibitem{pythia} T. Sjostrand, Comp. Phys. Commun. 
{\bf 89}, 74 (1994);
 S. Mrenna, Comp. Phys. Commun. 
{\bf 101}, 232 (1997).


\bibitem{cteq3}  
R. Brock {\sl et al.}, Rev. Mod. Phys. {\bf 67}, 157 (1995).

\bibitem{comphep} A. Pukhov {\sl et al.}, 
hep-ph/9908288.

\bibitem{beena}
 W. Beenakker {\sl et al.}, 
Nucl. Phys. {\bf B515}, 3 (1998)

\bibitem{olivier} B. Olivier, Ph. D. Thesis, University of Paris VI (2001)
(unpublished).


\bibitem{Bert} 
I. Bertram {\sl et al.}, 
Fermilab-TM-2104 (2000).

\bibitem{djouadi-stau} 
W.~Porod and T.~Wohrmann,
Phys. Rev. D {\bf 55}, 2907 (1997);
A.~Datta, M.~Guchait and K.~K.~Jeong,
Int. J. Mod. Phys.\ A {\bf 14}, 2239 (1999);
A.~Djouadi and Y.~Mambrini,
Phys.\ Rev.\ D {\bf 63}, 115005 (2001).    


\bibitem{lep2-stop}
LEP SUSY Working Group, ALEPH, DELPHI, L3 and OPAL Collaborations, 
LEPSUSYWG/01-02.1 (2001), 
{\tt http://lepsusy.web.cern.ch/lepsusy/}.




\end{references}
\end{document}